\documentclass[twocolumn]{aastex701}

\usepackage{hyperref}
\usepackage{amsmath}
\usepackage{caption}
\usepackage{subcaption}

\newcommand{\Msun}{M$_\odot$} 
\newcommand{\Mstar}{M$_\star$}
\newcommand{\Mbh}{$M_\textrm{BH}$} 
\newcommand{\mesa}{\texttt{MESA}} 

\newif\ifref
\reftrue
\reffalse
\definecolor{darkred}{rgb}{0.75, 0, 0}
\newcommand{\mb}[1]{\ifref\textcolor{darkred}{#1}\else #1\fi}

\begin{document}

\title{Evolutionary Tracks and Spectral Properties of Quasi-stars and \\Their Correlation with Little Red Dots}

\accepted{2026, Jan. 6}

\author[orcid=0009-0007-4394-3366, gname='Andy', sname='Santarelli']{Andrew D. Santarelli}
\affiliation{Department of Astronomy, Yale University}
\email[show]{andy.santarelli@yale.edu}  

\author[orcid=0000-0002-5794-4286,gname=Ebraheem,sname=Farag]{Ebraheem Farag}
\affiliation{Department of Astronomy, Yale University}
\email{ebraheem.farag@yale.edu}

\author[orcid=0000-0003-4456-4863,gname='Earl', sname='Bellinger']{Earl P. Bellinger} 
\affiliation{Department of Astronomy, Yale University}
\email{earl.bellinger@yale.edu}

\author[orcid=0000-0002-5554-8896,gname='Priyamvada', sname='Natarajan']{Priyamvada Natarajan} 
\affiliation{Department of Astronomy, Yale University}
\affiliation{Department of Physics, Yale University}
\affiliation{Yale Center for the Invisible Universe}
\affiliation{Black Hole Initiative, Harvard University}
\email{priyamvada.natarajan@yale.edu}

\author[orcid=0000-0003-3997-5705,gname=Rohan, sname=Naidu]{Rohan P. Naidu}
\affiliation{Kavli Institute for Astrophysics and Space Research, MIT}
\email{rnaidu@mit.edu}

\author[orcid=0009-0002-5773-3531,gname=Claire,sname=Campbell]{Claire B. Campbell}
\affiliation{Department of Physics, Illinois State University}
\email{cbcamp1@ilstu.edu}

\author[orcid=0009-0006-2122-5606,gname='Matt', sname='Caplan']{Matthew E. Caplan} 
\affiliation{Department of Physics, Illinois State University}
\email{mecapl1@ilstu.edu}
\affiliation{Department of Physics, University of Illinois
Urbana-Champaign}

\begin{abstract}
JWST has revealed a population of red, compact, high-redshift (${z\sim3-10}$) objects referred to as ``Little Red Dots'' (LRDs). These objects exhibit unusual spectral features reminiscent of stellar spectra with blackbody-like SEDs, large hydrogen Balmer breaks, Balmer line absorption, and classical stellar absorption features such as calcium H\&K and the calcium triplet. Following the recent suggestion that these may be actively accreting direct-collapse black holes in the process of assembly, i.e. quasi-stars, we present evolutionary models of quasi-stars using our recently released, publicly available MESA-QUEST modeling framework. We compute a grid of models spanning a range of black hole masses and predict the luminosities, temperatures, surface gravities, and lifetimes of these objects. We find that these models lie along a Hayashi track once they hit their ``late-stage'' which constitutes the majority of their lives ($\sim 20$~Myr). We present scaling relations for estimating the mass of a quasi-star as a function of the bolometric luminosity, as well as the bolometric luminosity as a function of the effective temperature for the Hayashi track. The short lifetimes in tandem with the observed number density of LRDs imply the possibility that every supermassive black hole detected at these epochs was once a quasi-star. We compare synthetic spectra of our quasi-star models to observations of LRDs, and show that these models are broadly capable of reproducing the continuum spectra of observed LRDs. These results indicate that quasi-stars are promising candidates for the origin of supermassive black holes via direct collapse in the early universe. 

\end{abstract}



\section{Introduction} \label{sec:intro}

The rapid emergence of supermassive black holes (SMBHs) within the first $\sim700$~Myr of cosmic history remains a major puzzle. 
Observations indicating black hole-to-stellar mass ratios up to $\sim10\%$, significantly higher than those observed locally \citep{Harikane_2023, Jones2025LRD, maiolino2025, Yue_2024}, suggest that the theoretically proposed ``direct collapse'' pathway for forming massive initial black hole seeds may have been in operation in the early Universe \citep{Whalen2023, Natarajan2024, coughlin2024}. 
In this scenario, a heavy seed black hole of order $10^4$~\Msun\ forms from the gravitational collapse of low metallicity gas in early-universe dense pre-galactic gas clouds in various cosmic settings, which then accretes at the Eddington limit until it fully forms a SMBH \citep{Haehnelt1993, loeb1994, eisenstein1995, Madau_2001, bromm_loeb2003, lodato2006, begelman2006, Natarajan_2017} or from the rapid amplification of a light initial seed that can grow at super-Eddington rates \citep{Devecchi_2009,Mayer2010,alexander2014}. 

The evolution of direct collapse heavy seed formation is expected to proceed in the form of a quasi-star \citep{lodato2006, Begelman_2008}. These are formed when the core of either a supermassive star or a pre-galactic gas cloud directly collapses initially into a stellar mass black hole around which the remaining gas is assembled as an envelope. This envelope around the growing black hole seed is then supported by the radiation pressure produced from accretion onto the black hole, which can  rapidly accrete at super-Eddington rates due to the efficient transport of energy via convection and jets, allowing the central black hole to accrete at roughly the Eddington limit of the entire quasi-star \citep{begelman2006, Begelman_2008, begelman2010, Ball2012MNRAS, Fiacconi2015, Campbell_2025}. 

Among the high-redshift, actively-accreting black hole observations by JWST are a subset of candidate active galactic nuclei (AGN) that have come to be known as ``Little Red Dots" (LRDs) -- high-redshift, morphologically compact, ultra-luminous objects with unique spectral characteristics \citep{matthee-LRD, Greene_2024,labbe25, akins2024, Juodzbalis2024, Kokorev_2024, kocevski2025LRD, hviding2025, ji2025, taylor2025}. These objects exhibit an unusually strong Balmer break as well as an extended, flat spectrum in the near-infrared (NIR). The combination of these features, in addition to a blue UV slope, results in a unique ``V"-shaped spectral energy distribution \citep[SED,][]{Barro_2024, killi-LRD, deGraaff2025, kocevski2025LRD, labbe2023LRD}. 

Currently, there is no complete explanation for what these LRDs are composed of and hence their unique signatures. The relative contributions of emitted flux originating from star formation versus accretion onto a central BH are debated. \mb{\cite{nandal2025} suggest that they can be explained by supermassive stars due to their high luminosities and low photospheric temperatures. \cite{zwick2025lrd} add on to this and suggest the supermassive star would be embedded within a self-gravitating accretion disc formed from gas-rich galaxy mergers in order to explain the two-blackbody, V-shaped continuum.}
Another explanation put forth by several studies suggests that LRDs are black holes embedded in dense gas, with \citet{Naidu2025} dubbing their model a ``black hole star'' \citep[e.g.,][]{deGraaff2025, Inayoshi_2025, ji2025, kido2025, liu2025, taylor2025}.
This is motivated by their nearly blackbody-like SEDs, prominent hydrogen Balmer breaks, Balmer line absorption, and classical stellar absorption features such as Ca~H\&K and the Ca~II triplet, which are consistent with emission from an optically thick, thermally emitting envelope surrounding an accreting black hole.
\citet{Jeon_2025} further proposed that LRDs could be remnants of direct collapse, consistent with the quasi-star scenario \citep{begelman2006, Begelman_2008, begelman2010}, and \citet{begelman2025littlereddotslatestage} have argued that they would most likely represent the ``late stage'' of this process where the BH has accreted at least 10\% the total mass of the system. 

Thus far, there has been no comparison between quasi-star evolutionary models and JWST's LRDs. In this work, we present simulations of quasi-star evolution using \mesa\ \citep{Paxton2011, Paxton2013, Paxton2015, Paxton2018, Paxton2019, Jermyn2023} and the MESA-QUEST modeling framework \citep{Campbell_2025, santarelli_2025}. 
We extract spectral predictions to compare with recent JWST observations of three LRDs: ``The Cliff'' \citep{deGraaff2025}, MoM-BH*-1 \citep{Naidu2025}, and UNCOVER-45924 \citep{labbe2024unambiguousagnbalmerbreak}. We compare our predictions with observations for both a quasi-star by itself and a quasi-star embedded within a star-forming galaxy, akin to the scenario predicted by \cite{Natarajan_2017}.
We also consider simple corrections for dust attenuation, but find little evidence for it, in line with recent findings that dust attenuation in LRDs may be subtle or negligible \citep{Casey_2025, deGraaff2025, Naidu2025, Setton_2025, Xiao2025}. 
We find that our fiducial $10^6$~\Msun\ quasi-star model broadly reproduces the continuum spectra of these LRDs, thereby lending evidence that SMBHs may assemble by passing through the direct-collapse quasi-star phase. 

The outline of our paper is as follows: in Section~\ref{sec:mesa}, we provide details of how we extract spectra from quasi-star models in MESA-QUEST. We present the results from models and comparisons with observational data of three JWST LRDs in Section~\ref{sec:results}, and close with a final discussion and implications of these results.

\section{Methods: extracting quasi-star spectra from MESA-QUEST} \label{sec:mesa}

\begin{figure*}
    \centering
    \includegraphics[width=0.9\linewidth]{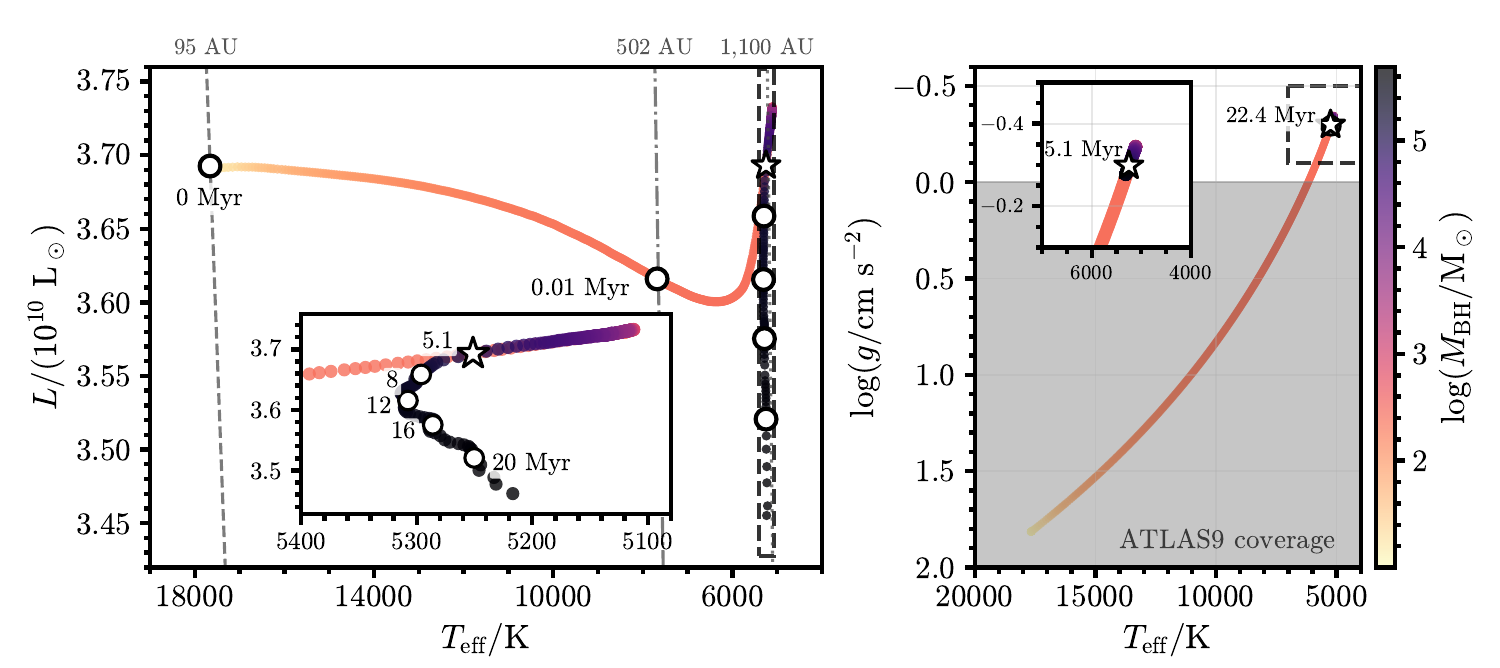}
    \caption{Evolution of a theoretical quasi-star model in Hertzsprung-Russel (left) and Kiel (right) diagrams with initial parameters $M_\mathrm{BH}=10$~\Msun, \Mstar$=10^6$~\Msun, and $Z=0$. After an initial contraction phase lasting $\sim0.01$~Myr, the model spends the remainder of its lifetime ($\sim$20~Myr) at $\sim 3.6\cdot 10^{10}~\textrm{L}_\odot$ and $\sim5250$~K before becoming a $\sim10^6$~\Msun~SMBH. Black hole mass is shown with a color gradient. Lines of constant radius are shown in gray, labeled above the top axis. Note that while these lines appear vertical, this is simply due to the small luminosity range. The gray region in the Kiel diagram represents the log($g$)--$T_\textrm{eff}$ space covered by the ATLAS9 atmosphere models. The star represents the beginning of the quasi-star's late-stage, approximately where the luminosity peaks on the HR diagram and where it spends the majority of its life.} 
    \label{fig:hr}
\end{figure*}

In this section, we briefly describe our quasi-star \mesa\ models. 
We use our previously developed MESA-QUEST framework, which includes several newly implemented upgrades: flexible inner boundary conditions, accretion schemes, and improvements that enable modeling higher total masses than previously studied. 
The following section briefly outlines these methods; more details can be found in \cite{santarelli_2025} and \cite{Campbell_2025}. 
We note that all \mesa\ work directories are publicly available on Zenodo with a living version available via the MESA-QUEST Github repository.\footnote{Zenodo:~\dataset[10.5281/zenodo.17992486]{https://doi.org/10.5281/zenodo.17992486} \\
GitHub:~\href{https://github.com/andysantarelli/MESA-QUEST}{https://github.com/andysantarelli/MESA-QUEST}}

\subsection{Quasi-star Modeling} \label{sec:bh}

A quasi-star is modeled as a black hole enveloped within a star via an adjustment to the inner boundary conditions of the stellar structure equations such that:
\begin{align}
    m(r_0=R_\mathrm{in}) &= M_\mathrm{BH} \\
    l(r_0=R_\mathrm{in}) &= L_\mathrm{BH}
\end{align}

\noindent where $r_0$ is the innermost grid point, $R_\mathrm{in}$ is the updated inner boundary, and $M_\mathrm{BH}$ and $L_\mathrm{BH}$ are the black hole mass and accretion luminosity respectively. In this work, we use the ``saturated-convection'' radius, the boundary within which material is slowly drifting inward while convection transports energy outward at the highest possible rate \citep{coughlin2024}. Details on the calculation of this boundary can be found in \cite{coughlin2024}, with further implementation details in \cite{santarelli_2025} and \cite{Campbell_2025}. For the accretion luminosity, and thus the accretion rate, we use the Eddington limit of the entire object such that

\begin{equation}\label{eq:alphaEdd}
    L_0 = \alpha L_\mathrm{E} = \alpha\, 4\pi \,\frac{c}{\kappa}\, GM_\star 
\end{equation}

\begin{equation}
    \dot M_{\rm{BH}} = \frac{1-\epsilon}{\epsilon}\, \frac{4\pi}{\kappa c}\,\alpha\, G M_\star.
\end{equation}

\noindent Previous works modeling the accretion and resulting luminosity report only a factor of $<2$ deviation from the object's Eddington limit throughout the quasi-star's lifetime \citep{Ball2012MNRAS, Campbell_2025}. The dimensionless scaling factor $\alpha$ serves as a proxy for a range of physical effects that enhance or suppress accretion such as rotation, magnetic fields, photon-trapping, etc., which we plan to implement and explore more thoroughly in future work. Here we adopt a simple scaling factor $\alpha$ and use $\alpha=1$ as a first approximation. 

Fig.~\ref{fig:hr} shows a detailed single evolutionary track for a $10^6$~\Msun\ total mass quasi-star with an initial BH mass of $10$~\Msun\ and metallicity $Z=0$. The timescale of the initial tail is short and comparable to the Kelvin-Helmholtz timescale,
\begin{align}
    t_\mathrm{KH} \simeq 0.04~\mathrm{Myr} &\left(\frac{M}{10^6~M_\odot}\right)^2\notag \left(\frac{R}{2.4\cdot10^4~R_\odot}\right)^{-1}
    \\&\qquad\left(\frac{L}{3.7\cdot10^{10}~L_\odot}\right)^{-1}
\end{align}

\noindent after which the quasi-star ``settles" as the black hole mass reaches $\sim10^3$~\Msun. \mb{It is unclear whether this initial phase is physical or a result of the initialization of the model.} The inset plot shows the majority of the quasi-star's life. 

The final age of this model is $\sim20$~Myr, \mb{i.e., the timescale for which a 10~\Msun\ black hole accretes at the Eddington limit of the entire object until $M_\mathrm{BH}/M_\star \approx 0.5$}. The actual lifetimes may be longer as our models do not currently indicate what happens once the inner boundary condition exceeds the total radius, \mb{and it has been suggested that the envelope would be ejected once the surface temperature cools to $\lesssim 4000$~K \citep{Begelman_2008, Ball2012MNRAS}. It is unclear when or if this would occur after our model is terminated.} If we assume that BH growth continues at the Eddington rate of the entire object until the envelope is completely consumed, we expect a total lifetime of $30-40$~Myr. 

\begin{figure*}
    \centering
    \begin{subfigure}[b]{0.49\textwidth}
        \centering
        \includegraphics[width=\textwidth]{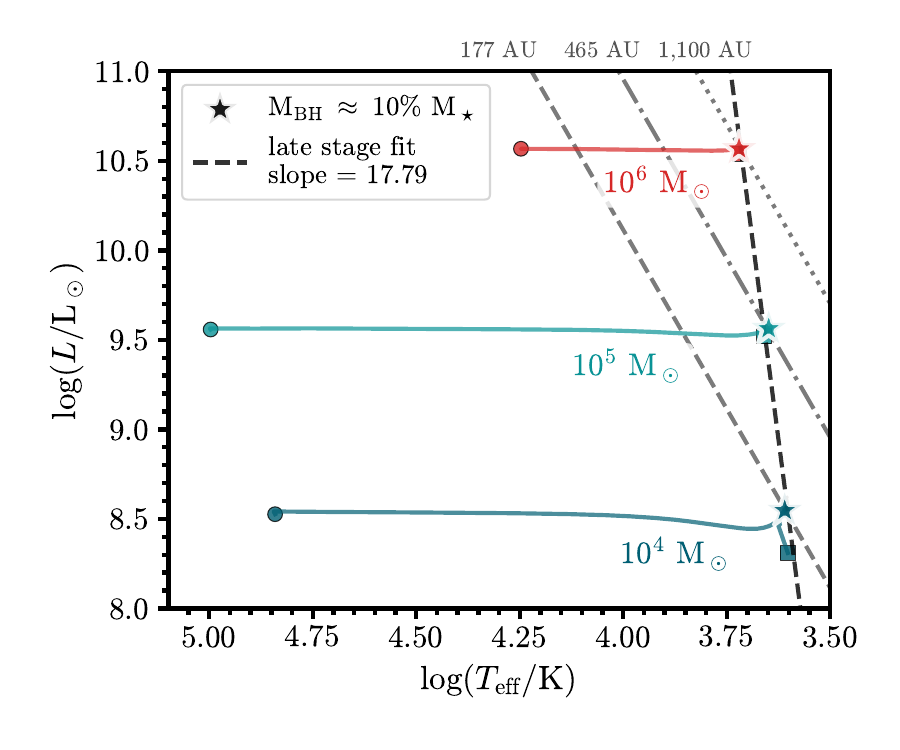}
    \end{subfigure}
    \hfill
    \begin{subfigure}[b]{0.49\textwidth}
        \centering
        \includegraphics[width=0.95\textwidth]{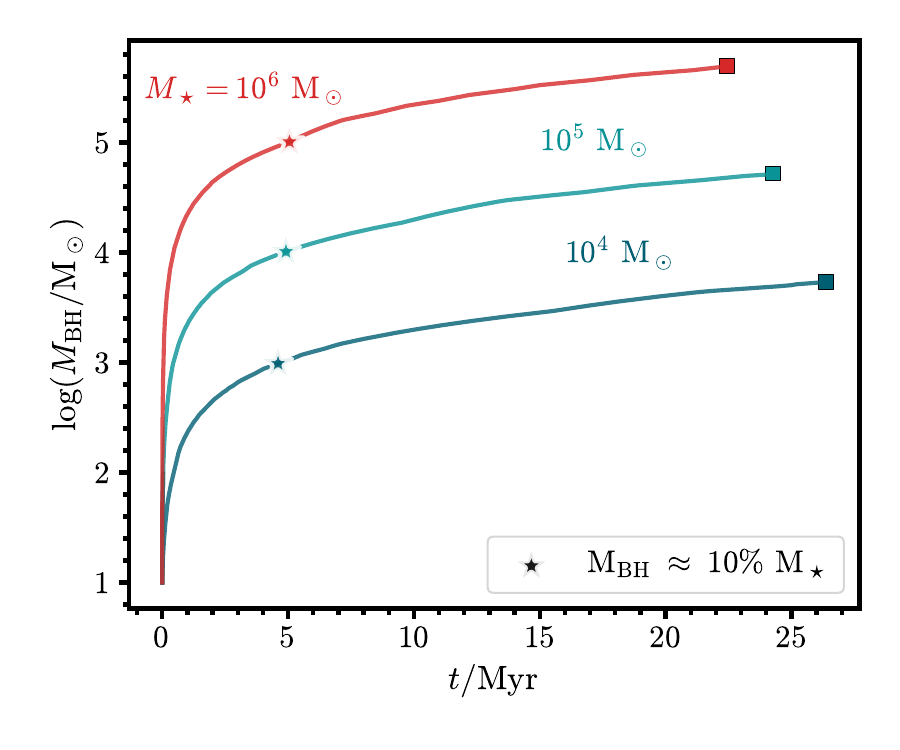}
    \end{subfigure}
    
    \caption{Evolution of three quasi-stars with masses $10^4$~\Msun\ (blue), $10^5$~\Msun\ (green), and $10^6$~\Msun\ (red). 
    A Hertzsprung-Russel diagram is shown on the left, and the black hole mass growth with age on the right. 
    The start and end points in the HR diagram are marked with circles and squares, respectively. 
    The beginning of the late-stage period tracing a Hayashi track is shown with star symbols, and the fit is depicted with a black dashed line. 
    Lines of constant radius are shown in gray and labeled above the top axis.}
    \label{fig:hr-age}
\end{figure*}

We show evolutionary tracks for multiple masses in Fig.~\ref{fig:hr-age}, including a line marking where the quasi-star models enter their ``late stage'' i.e. \Mbh/\Mstar$=0.1$. Due to the large accretion luminosity and high opacity, the envelope becomes fully convective and the model evolves to low effective temperatures and hence occupies the Hayashi track --- the hydrostatic limit for such structures --- for most of its lifetime, akin to fully convective pre-main-sequence stars and red giants with deep convective envelopes. 
The fit for their Hayashi track can be reasonably well represented by 

\begin{equation}
    \mathrm{log}(L/\mathrm{L_\odot}) = 18~\mathrm{log}(T_\mathrm{eff}/\mathrm{K}) - 55.
\end{equation}

\noindent One can in principle also obtain an estimate for the mass of an observed quasi-star based on its luminosity through Eq.~\ref{eq:alphaEdd} via 

\begin{align}
    M_\star \simeq 10^{6}\,M_\odot\,
    \left(\frac{L}{3.7\cdot 10^{10}\,\rm L_\odot}\right)
\end{align}
where for convenience we have dropped the dependency on $\kappa$ and $\alpha$, though we note that their variation can have a large influence on the accretion physics and thus the luminosity itself \citep{santarelli_2025}. 

\begin{figure*}
    \centering
    \includegraphics[width=\linewidth]{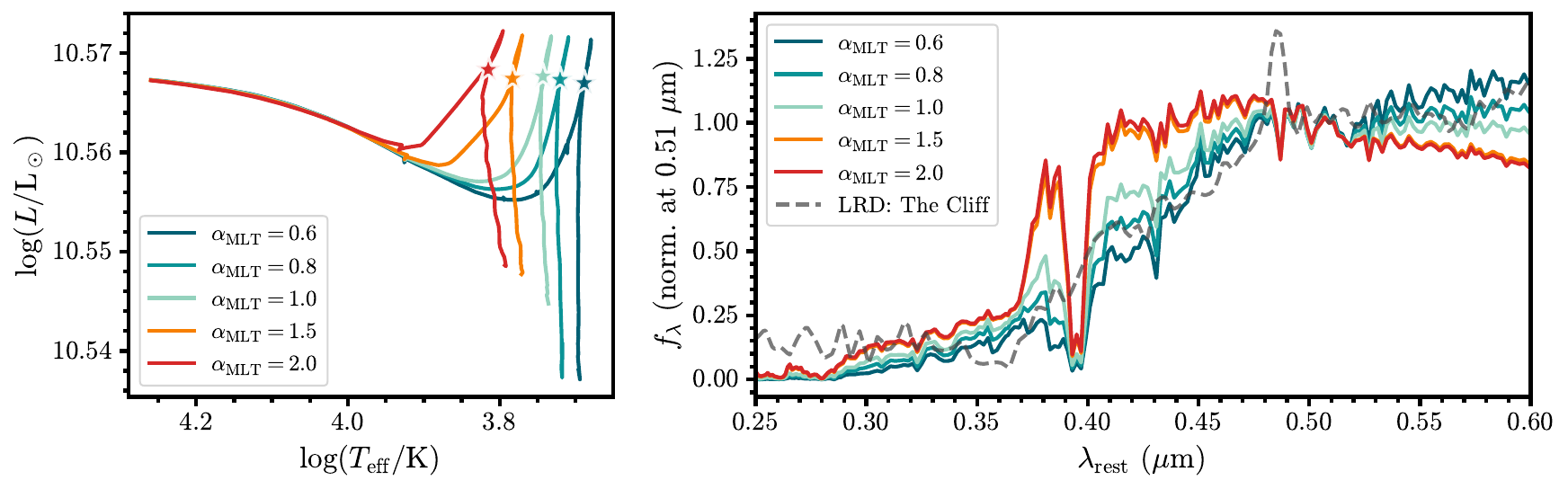}
    \caption{\mb{A Hertzsprung-Russel diagram showing the evolution of $10^6$~\Msun\ quasi-stars using a range of mixing length parameters $\alpha_\mathrm{MLT}=0.6$ to $2.0$ with the beginning of their late-stage (i.e. where $M_\mathrm{BH}/\mathrm{M}_\odot = 0.1$ marked by a star is shown on the left. SEDs corresponding with these models at the late-stage are shown on the right. We show the SED for the LRD ``The Cliff'' (gray, dashed) for reference, more detailed comparisons and visualizations can be found in Fig.~\ref{fig:lrd}.} \label{fig:mlt}}
\end{figure*}

\mb{Within mixing length theory, the mixing length parameter $\alpha_\mathrm{MLT}$ is a dimensionless parameter that defines the distance that a parcel of convective material can travel in terms of the pressure scale height \citep{bohm-vitense1958, cox1968}. While \mesa\ uses a default value of $2.0$, $\alpha_\mathrm{MLT}$ is a free parameter of order unity that should vary across models. We show in the left plot of Fig.~\ref{fig:mlt} the impact that it has on our models: decreasing $\alpha_\mathrm{MLT}$ results in an increase in radius and a lower $T_\mathrm{eff}$, and thus a reddening of the SED. For this work, we choose $\alpha_\mathrm{MLT}=0.8$ and $1.0$ for our $10^6$~\Msun\ total mass models as they best fit the SEDs of the selected LRDs. While models with masses of $\lesssim10^5$~\Msun\ have similar effective temperatures with higher values of $\alpha_\mathrm{MLT}$, they do not have the bolometric luminosities seen in most LRDs. Further details on extraction of the SEDs from the models as well as the LRD comparisons can be found in Sections~\ref{sec:sed} and \ref{sec:results}.}

\mb{A detailed analysis of the impact of our modeling assumptions (e.g. accretion rate, boundary conditions, surface winds) can be found in \cite{santarelli_2025} with additional comparison of a similar model to previous analytic models in \cite{hassan}. \cite{santarelli_2025} shows that changes to the accretion rate, particularly for accretion stifling cases, may result in smaller radii and thus hotter surface temperatures. The inclusion of surface winds is shown to have a large impact on the stability of the quasi-star and thus the final black hole mass. Both the black hole accretion physics and the offset to the wind-driven mass loss from envelope accretion remain uncertain and will be studied in future work.} 

\mb{We have explored the inclusion of sparse metals within the envelope as well as alternate opacity tables. Changes to either of these parameters results in no qualitative difference in the evolution or emergent spectra discussed in the following section. The alternate opacity tables have negligible impact due to the efficient convection in the envelope. For models with increased metallicity, the similarities in spectra are likely due to the limited parameter space covered by the adopted atmospheric grids (see Sec.~\ref{sec:sed} and Fig.~\ref{fig:kiel}).}

\subsection{Extracting Synthetic Spectral-Energy Distributions} \label{sec:sed}

In order to extract the synthetic spectra of quasi-stars, we use the \mesa\ \texttt{colors} module, developed by \cite{COLORS}. This is a built-in tool within MESA that can be used to generate synthetic spectral-energy distributions (SEDs) by sampling the nearest grid point in a stellar atmosphere model grid using the stellar surface conditions output by \mesa\ (i.e.~$T_\textrm{eff}$, log($g$), and metallicity $Z$). Furthermore, colors can then convolve these synthetic SEDs with filter transmission curves from astronomical surveys including $Gaia$ and $JWST$, the latter of which we use in this work. 

\begin{figure}
    \centering
    \includegraphics[width=\linewidth]{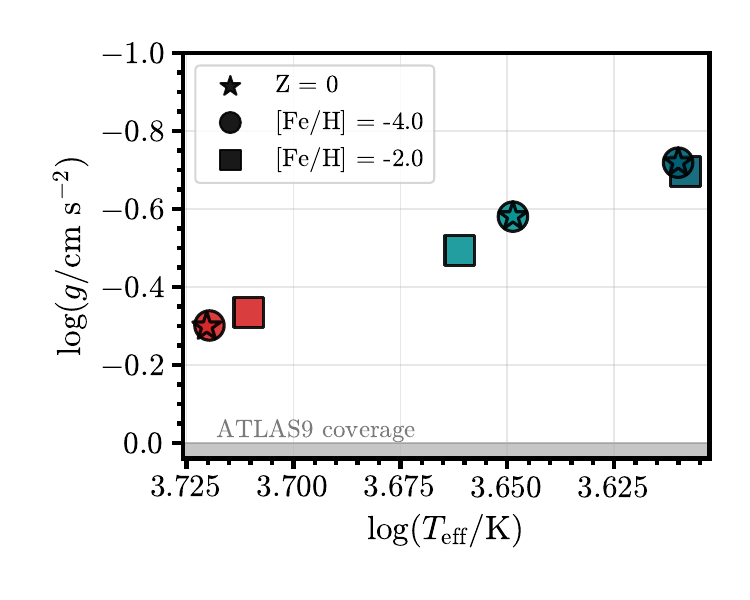}
    \caption{A set of quasi-star models with total masses $10^4$~\Msun\ (blue), $10^5$~\Msun\ (green), and $10^6$~\Msun\ (red) when $M_\textrm{BH} \approx 0.1 M_\star$, which is the point at which the quasi-star has reached what we consider the ``late-stage'' where it remains predominantly static within the HR diagram. The gray shaded region indicates the space covered by the ATLAS9 grids. Additionally, these grids span $-5 \leq \textrm{[Fe/H]} \leq 1.5$.}
    \label{fig:kiel} 
\end{figure}

We use the ATLAS9 atmospheric model grid \citep{atlas9}, the data and interpolation for which is provided by default in the \texttt{colors} module. However, it should be noted that this grid does not span the full parameter space to cover that of quasi-stars at all points in their lifetimes. Fig.~\ref{fig:kiel} shows a Kiel diagram containing points for various quasi-star models, \rm{all with $\alpha_\mathrm{MLT}=0.8$.} The small gray region shows the log($g$) -- $T_\textrm{eff}$ space covered by the ATLAS9 grid. Although all model temperatures lie firmly within the grid range, the models lie outside of the log($g$) range. Furthermore, the lowest metallicity provided is $\textrm{[Fe/H]} = -5$. However, comparison of the low and zero metallicity models shows very little difference in the final SEDs \mb{and log($g$) is off of the grid by $\sim0.3$ in log-space for the $10^6$~\Msun\ model with $\alpha_\mathrm{MLT}=0.8$, and $\sim0.2$ when $\alpha_\mathrm{MLT}=1.0$.} Therefore we use a $10^6$~\Msun, $Z=0$ model for all SEDs in this work. 

A few caveats arise due to our adoption of the ATLAS9 atmosphere models. 
First, these models are computed in local thermodynamic equilibrium (LTE) and therefore lack the physics necessary to produce all of the emission lines required for rigorous analysis. 
Additionally, a solar-scaled abundance mixture is assumed, and thus absorption features that would not otherwise be present in metal-free atmospheres are seen. 
This appears primarily in our models on the H and K singly-ionized calcium (CaII) lines at 3968~\AA\ and 3934~\AA\ respectively. 
While these strong absorption features are present in the adopted ATLAS9 atmosphere models, they can safely be ignored in this work, as there is little to no Ca present in the atmospheres of our quasi-star models and are expected to be far less abundant in the composition of the gas at the high redshifts studied in this work \mb{than in the solar-scaled abundances in the atmosphere models.} It is, however, conceivable that quasi-stars may have some enrichment via accretion onto the envelope from a host galaxy that has undergone some star formation.
We emphasize again that our comparisons are between the continuum features of the models and observations and serve as a pilot study. 
The modeling and implementation of more detailed atmosphere models will be the subject of future study. 

\subsection{Host Galaxy \& Dust Corrections} \label{sec:galaxy}

Multiple explanations for the observed LRD spectra have been proposed. Leading theories are that these are compact, luminous objects (such as a quasi-star or black hole star) with additional components in the emission deriving from host galaxies or dust, although dust is presently disfavored \citep{deGraaff2025, labbe2024unambiguousagnbalmerbreak, Naidu2025, Setton_2025}. Observations are presently inconclusive with regard to the issue of the contribution to the flux from the stellar component of the host galaxies. Additionally, dust properties may vary between sources and with redshift, so we remain agnostic when exploring the impact of the stellar population of a host galaxy and dust to our models.

In order to combine our quasi-star model with that of its host galaxy, we use spectral data from the Dawn JWST Archive \citep[DJA,][]{dja1, dja2, dja3}. We create semi-synthetic models by simply combining our theoretical model with an observed star-forming galaxy at a redshift similar to the observed LRD. 
The hosts we have selected (JADES-27368, MoM-UDS-948311, and UNCOVER-24996) primarily contribute to the UV spectrum where they dominate, and are found at sky positions and redshifts similar to their respective LRDs ($z\simeq 3.5,\ 7.7,$ and $4.4$ respectively) \citep{JADES2025, naidu2024, Naidu2025}. 
In the quasi-star scenario, a host similar to those selected would serve as the ionizing source that prevents fragmentation due to molecular hydrogen, thus allowing the formation of the quasi-star \citep{Agarwal+2013, Agarwal2016, begelman2006, Natarajan_2017, Naidu2025, Regan+2017, Shang+2010, Visbal_2018}. 

We implement dust corrections to our models using the simplified methods of \cite{Salim2020-dust} via 
\begin{equation}
    m_\lambda = m_{\lambda,0} + A_\lambda
\end{equation}
where $m_{\lambda,0}$ is the dust-free model (i.e. the quasi-star on its own) and $A_\lambda$ is the dust curve itself. Although there are several components that go into calculating $A_\lambda$, we use a simple case based only on the UV-optical slope $S$ and a small UV-bump with strength $B$. In this work, we use optical depth $A_V=1.6$ and bump strength $B=0.1$. It should be noted that the effects of this bump strength are minimal and can be safely ignored if necessary, and that $A_V$ values this high have been ruled out for several LRDs \citep{Setton_2025}. 

\section{Results} \label{sec:results}

\begin{figure*}
    \centering
    \begin{subfigure}[b]{0.49\textwidth}
        \centering
        \includegraphics[width=\textwidth]{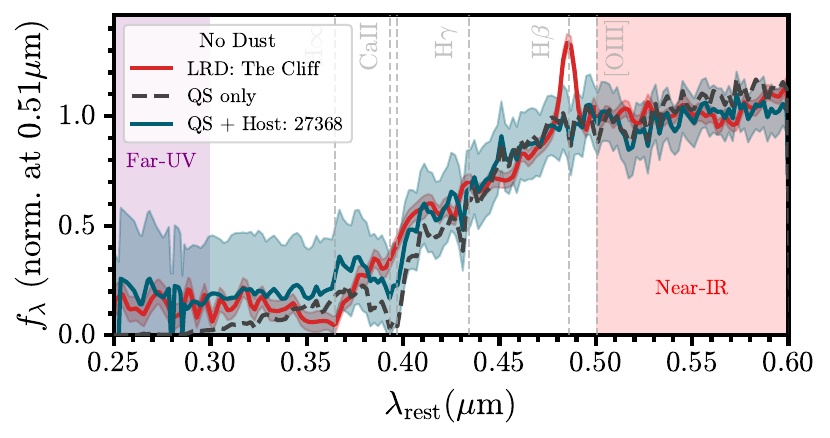}
    \end{subfigure}
    \hfill
    \begin{subfigure}[b]{0.49\textwidth}
        \centering
        \includegraphics[width=\textwidth]{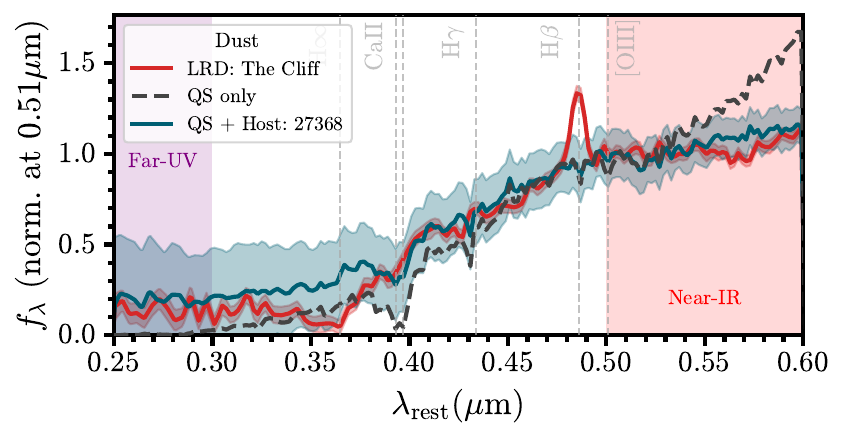}
    \end{subfigure}
    \begin{subfigure}[b]{0.49\textwidth}
        \centering
        \includegraphics[width=\textwidth]{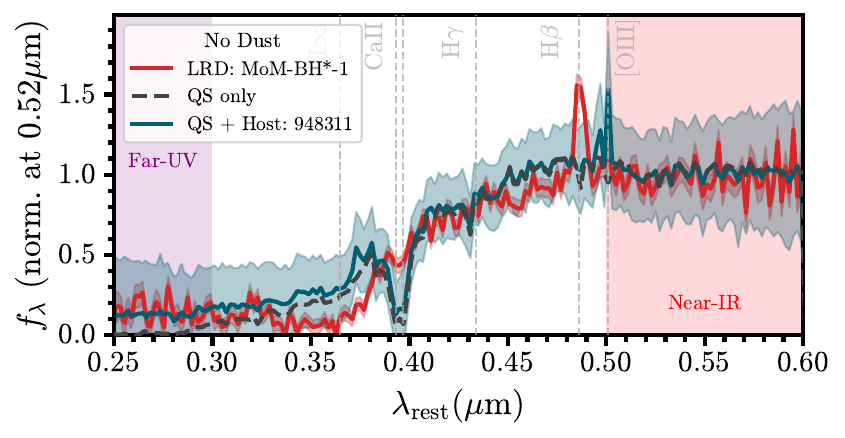}
    \end{subfigure}
    \hfill
    \begin{subfigure}[b]{0.49\textwidth}
        \centering
        \includegraphics[width=\textwidth]{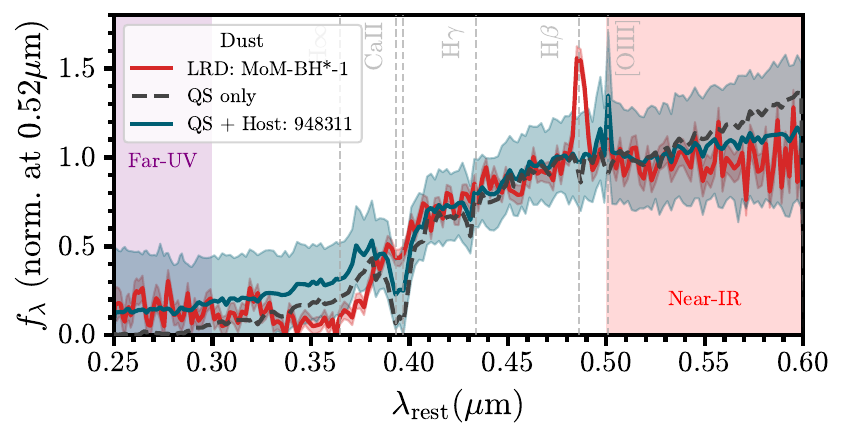}
     \end{subfigure}
     \hfill
     \begin{subfigure}[b]{0.49\textwidth}
        \centering
        \includegraphics[width=\textwidth]{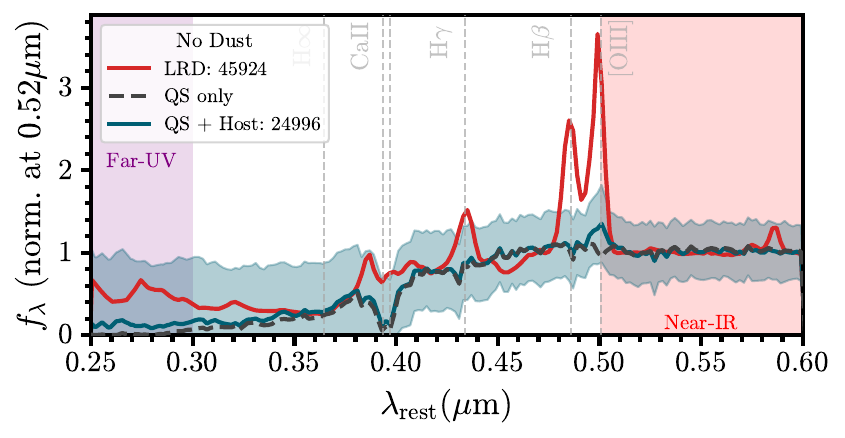}
    \end{subfigure}
    \hfill
    \begin{subfigure}[b]{0.49\textwidth}
        \centering
        \includegraphics[width=\textwidth]{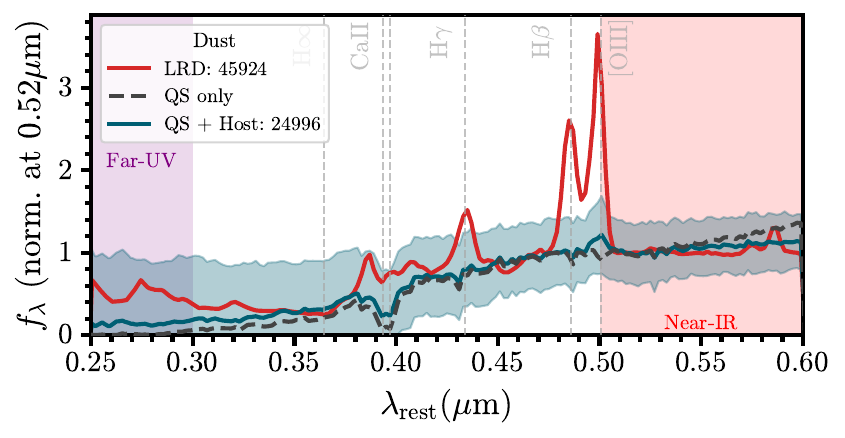}
    \end{subfigure}

     \caption{Predicted SEDs of a $10^6$~\Msun\ total mass, late-stage quasi-stars with $\alpha_\mathrm{MLT}=0.8$ (top) and $1.0$ (middle and bottom), both on their own (black dashed line) and embedded in a host galaxy (solid blue line) compared with JWST observations of LRDs (red) that includes the emission from the host stellar component with masses $\sim10^{9.5}$~\Msun. \mb{The top row shows comparisons with ``The Cliff''} \citep{deGraaff2025}, the middle row with MoM-BH*-1 \citep{Naidu2025}, and the bottom row with UNCOVER-45924 \citep{labbe2024unambiguousagnbalmerbreak}. We show models with (left) and without (right) dust. Semi-synthetic models of quasi-stars embedded in a host galaxy are shown in blue, using hosts properties inferred for JADES-27368, MoM-UDS-948311, and UNCOVER-24996 respectively \citep{naidu2024}. The faded colors surrounding the lines represent $1\sigma$ uncertainties in JWST measurements. Gray dashed lines indicate key wavelengths: the Balmer break (H$\infty$), H$\gamma$, H$\beta$, [OIII], and both CaII lines. Discrepancies in emission lines likely arise from the use of LTE atmospheres at nonzero metallicity, particularly in the H$\beta$ emission and CaII absorption lines - these are discussed in Sec.~\ref{sec:sed}. Shaded regions indicate the far-UV and NIR approach, highlighting a few of the well fit continuum regions discussed in Sec.~\ref{sec:results}.}
    \label{fig:lrd}

\end{figure*}

Here we compare the predicted spectra of a fiducial $10^6$~\Msun\ quasi-star model with JWST observations of \mb{three} LRDs: \mb{``The Cliff''}, MoM-BH*-1, and UNCOVER-45924 \citep{deGraaff2025, labbe2024unambiguousagnbalmerbreak, Naidu2025}. 
We present the former two as examples with extreme Balmer breaks and only small contributions from a host galaxy, and the latter as an example of a more ``typical'' LRD with significant host contributions. 
To facilitate this comparison, we normalize the flux at wavelengths corresponding to continuum features, and correct the wavelengths in the model and observations for the redshift in order to compare in the rest frame. 

Although the quasi-star models were not specifically fit to any LRD, they qualitatively reproduce the main continuum features seen in all three presented here. 
In order to make this comparison more quantitative, and despite the aforementioned limitations of our adopted atmospheric models which prevent detailed comparisons of individual spectral features, we compare the slopes of the UV and NIR-approach as well as the strength of the Balmer break. We find general agreement, with a few caveats listed below. 

For each of our configurations (i.e. a quasi-star with and without a host, each with and without dust), we calculate the UV and NIR-approach slope parameters $\beta_\mathrm{blue},~\beta_\mathrm{red}$ (where $f_\lambda \propto \lambda^\beta$) by performing linear regression in log-log space over the rest-frame wavelength ranges 0.18--0.28~$\mu$m and 0.51--0.57~$\mu$m, respectively. We then do the same for each LRD and compare the results. We find that the comparisons without dust are in good agreement, particularly with when a host is included ($<0.6\sigma$ in the NIR, $<1.5\sigma$). All cases that include dust have significant discrepancies in the NIR, concordant with other studies finding minimal evidence of dust attenuation in LRDs \citep{Setton_2025}.

We calculate the Balmer break strengths \mb{of our models and the LRDs using a variation on the the revised index from \cite{Wang_2024}. We use the mean flux in the same blue wavelength window ($3620-3720$~\AA) and change the red window from the original $4000-4100$~\AA\ to $4100-4200$~\AA\ in order to avoid the CaII absorption features in our models discussed in Sec. \ref{sec:sed}. Note that this leads to slightly different values for the Balmer break strengths than reported in each LRD's original publication.} 
Our model yields a break strength of $3.67$, closely matching the UNCOVER LRD’s value of $3.52\pm0.05$. 
The value of the break in our model does not change significantly with the addition of a host galaxy, as the Balmer break window is dominated by the quasi-star flux. 
Although the far-UV portion of the standalone quasi-star is relatively weak, we obtain the characteristic LRD ``V"-shape with the addition of a host galaxy. 
The inclusion of dust in the model \mb{reduces their similarities to the observed LRDs in all cases, deviating significantly in the NIR in particular.} This is in agreement with recent claims that LRDs are mostly dust-free \citep{Casey_2025, Naidu2025, Setton_2025, Xiao2025}.

\mb{The sources ``The Cliff'' and MoM-BH*-1 have some of the most pronounced Balmer breaks ($7.67 \pm 2.18$ and $7.82 \pm 1.24$ respectively)} of any LRD discovered \citep{Naidu2025}. While this is a more pronounced break than in our model ($3.67$), as noted previously, this comparison may improve significantly with the adoption of more appropriate atmosphere models. Furthermore, this comparison is with a fiducial $10^{6}$~\Msun\ quasi-star model not specifically fit to this source; a higher-mass model may reproduce these features more closely. 
Based on Fig.~\ref{fig:hr} we can estimate that a quasi-star model containing a more massive black hole of \textit{a few}$\times 10^6$~\Msun\ will closely match the LRD's bolometric luminosity. 
These topics will be the subjects of future investigations. 

\section{Conclusions \& Discussion} \label{sec:conclusions}

We have presented new \mesa\ quasi-star evolutionary models generated with the MESA-QUEST framework and compared their synthetic spectra to JWST observations of LRDs.
Our results demonstrate that late-stage quasi-stars, with black holes comprising $\gtrsim$10\% of their total mass and photospheric temperatures near $\sim5250$~K, naturally produce the defining continuum features of LRDs: large Balmer breaks, flat or red near-infrared slopes, and weak ultraviolet emission. These properties emerge self-consistently from the quasi-star's radiation-supported envelope, and do not require an old stellar population or extreme dust reddening. \mb{While some of these features may also appear in supermassive stars, they are expected to have both order of magnitude lower luminosities and up to several thousand K greater temperatures than those found in late-stage quasi-stars of equivalent masses, as well as extremely short lifetimes in which they form and collapse within $\sim 1000$ years \citep{Hosokawa+2013, nandal2025}.}

The quasi-star phase of direct collapse formation of heavy seed black holes offers a plausible solution to the rapid emergence of supermassive black holes within the first few hundred million years of the Universe. 
\mb{With lifetimes of $\sim20-40$~Myr, quasi-stars would exist for roughly $2-4\%$ of the age of the universe at redshift $z \sim 5$. Combining this duty cycle with the observed LRD comoving density of $\sim 10^{-5}$~cMpc$^{-3}$ at similar redshifts \citep{Greene_2024}, we estimate that the total number density of objects that undergo an LRD phase is $2.5-5 \times 10^{-4}$~cMpc$^{-3}$. If quasi-stars have a median mass of $\sim10^6$~\Msun\ and subsequently accrete an additional factor of $\sim 100$ during their post-quasi-star evolution, the cumulative black hole mass density contributed by this population reaches $\sim 2.5-5 \times 10^4$~\Msun cMpc$^{-3}$. Optimistically, this value is only about a factor of two below the observed black hole mass density of $\sim 2.5 \times 10^5$~\Msun cMpc$^{-3}$ \citep{yu2002}, suggesting that} a population of short-lived, extremely luminous quasi-stars could account for a significant fraction of present-day supermassive black holes. Folding in lifetimes from our models along with predictions of progenitor cloud masses will allow us to predict if this is the case. Given the short lifetimes in our models, this can plausibly show that a significant fraction, or even \textit{all} SMBHs formed this way.

Our synthetic spectra are limited to LTE stellar-atmosphere approximations and therefore capture continuum but not emission-line physics; future non-LTE radiative-transfer models will be required to capture these additional features. Nevertheless, the present results establish quasi-stars as viable progenitors of both LRDs and the earliest supermassive black holes. If JWST's LRDs indeed trace this brief phase, we may be witnessing the birth of supermassive black holes at cosmic dawn.

\begin{acknowledgments}
The authors thank Niall J. Miller for his contributions to \mesa\ and the \texttt{colors} module.

Financial support for this publication comes from Cottrell Scholar Award \#CS-CSA-2023-139 sponsored by Research Corporation for Science Advancement. This work was supported by a grant from the Simons Foundation (MP-SCMPS-00001470) to MC. This research was supported in part by the National Science Foundation under Grant No. NSF PHY-1748958. P.N. acknowledges support from the Gordon and Betty Moore Foundation and the John Templeton Foundation that fund the Black Hole Initiative (BHI) at Harvard University where she serves as one of the PIs. This research is supported by the Yale Center for Astronomy and Astrophysics Prize Fellowship (EF).

Some of the data products presented herein were retrieved from the Dawn JWST Archive (DJA). DJA is an initiative of the Cosmic Dawn Center (DAWN), which is funded by the Danish National Research Foundation under grant DNRF140.

\mb{The JWST data used in this work is available via doi:\dataset[10.17909/8tdj-8n28]{\doi{10.17909/8tdj-8n28}}, doi:\dataset[10.17909/8k5c-xr27]{\doi{10.17909/8k5c-xr27}}, and doi:\dataset[10.17909/4xx0-zj76]{\doi{10.17909/4xx0-zj76}}.}
\end{acknowledgments}

\bibliography{refs, mesa, dja}{}
\bibliographystyle{aasjournalv7}

\end{document}